\documentclass[twocolumn,iop,numberedappendix,appendixfloats]{oja}

\usepackage[utf8]{inputenc}
\usepackage[british]{babel}

\usepackage{graphicx}	
\usepackage{amsmath}	
\usepackage{amssymb}	
\usepackage{subcaption}  
\usepackage{caption}
\usepackage{booktabs}

\usepackage{xcolor}
\definecolor{maroon}{cmyk}{0, 0.87, 0.68, 0.32}
\definecolor{halfgray}{gray}{0.55}
\definecolor{ipython_frame}{RGB}{207, 207, 207}
\definecolor{ipython_bg}{RGB}{247, 247, 247}
\definecolor{ipython_red}{RGB}{186, 33, 33}
\definecolor{ipython_green}{RGB}{0, 128, 0}
\definecolor{ipython_cyan}{RGB}{64, 128, 128}
\definecolor{ipython_purple}{RGB}{170, 34, 255}
\usepackage{float}
\usepackage{fontawesome}
\usepackage[colorlinks,allcolors=blue]{hyperref}
\usepackage{url}

\usepackage{listings}

\usepackage{array,multirow,graphicx}

\begin{document}

\journalinfo{The Open Journal of Astrophysics}
\submitted{submitted XXX; accepted YYY}

\title{The future of cosmological likelihood-based inference:\\ accelerated high-dimensional parameter estimation and model comparison\vspace{-10ex}}

\shorttitle{The future of likelihood-based cosmology}
\shortauthors{Piras et al.}

\author{Davide Piras$^{\star \dagger1,2}$, Alicja Polanska$^{\dagger3}$, Alessio Spurio Mancini$^{4,3}$, Matthew A. Price$^{3}$, Jason D. McEwen$^{3,5}$}

\affiliation{$^1$ Centre Universitaire d’Informatique, Université de Genève, 7 route de Drize, 1227 Genève, Switzerland}

\affiliation{$^2$ Département de Physique Théorique, Université de Genève, 24 quai Ernest Ansermet, 1211 Genève 4, Switzerland}

\affiliation{$^3$ Mullard Space Science Laboratory, University College London, Holmbury St. Mary, Dorking, Surrey, RH5 6NT, UK}

\affiliation{$^4$ Department of Physics, Royal Holloway, University of London, Egham Hill, Egham, UK}

\affiliation{$^5$Alan Turing Institute, London, NW1 2DB, UK}

\thanks{$^\star$ E-mail: \href{mailto:davide.piras@unige.ch}{davide.piras@unige.ch}}
\thanks{$^\dagger$ Joint first authors}

\date{\today}

\begin{abstract}
	We advocate for a new paradigm of cosmological likelihood-based inference, leveraging recent developments in machine learning and its underlying technology, to accelerate Bayesian inference in high-dimensional settings. Specifically, we combine (i) \textit{emulation}, where a machine learning model is trained to mimic cosmological observables, e.g.\ \texttt{CosmoPower-JAX}; (ii) \textit{differentiable and probabilistic programming}, e.g.\ \texttt{JAX} and \texttt{NumPyro}, respectively; (iii) \textit{scalable Markov chain Monte Carlo (MCMC) sampling} techniques that exploit gradients, e.g.\ Hamiltonian Monte Carlo; and (iv) \textit{decoupled and scalable Bayesian model selection} techniques that compute the Bayesian evidence purely from posterior samples, e.g.\ the learned harmonic mean implemented in \texttt{harmonic}.
	This paradigm allows us to carry out a complete Bayesian analysis, including both parameter estimation and model selection, in a fraction of the time of traditional approaches.
	First, we demonstrate the application of this paradigm on a simulated cosmic shear analysis for a Stage IV survey in 37- and 39-dimensional parameter spaces, comparing $\Lambda$CDM and a dynamical dark energy model ($w_0w_a$CDM).  We recover posterior contours and evidence estimates that are in excellent agreement with those computed by the traditional nested sampling approach while reducing the computational cost from 8 months on 48 CPU cores to 2 days on 12 GPUs.
	Second, we consider a joint analysis between three simulated next-generation surveys, each performing a 3x2pt analysis, resulting in 157- and 159-dimensional parameter spaces.
	Standard nested sampling techniques are simply unlikely to be feasible in this high-dimensional setting, requiring a projected 12 years of compute time on 48 CPU cores; on the other hand, the proposed approach only requires 8 days of compute time on 24 GPUs.
	All packages used in our analyses are publicly available.
\end{abstract}

\maketitle



\section{Introduction}

The evolution of cosmological likelihood-based data analysis is heading towards a high-dimensional future. Fueled by the acquisition of ever more constraining observational data thanks to ongoing and upcoming surveys like \textit{Euclid} \citep{Laureijs11}\footnote{\href{https://www.euclid-ec.org/}{https://www.euclid-ec.org/}}, the Dark Energy Spectroscopic Instrument \citep[DESI,][]{Levi19}\footnote{\href{https://www.desi.lbl.gov/}{https://www.desi.lbl.gov/}}, the Nancy Grace Roman Space Telescope \citep{Spergel15}\footnote{\href{https://roman.gsfc.nasa.gov/}{https://roman.gsfc.nasa.gov/}}, the Simons Observatory \citep{Ade19}\footnote{\href{https://simonsobservatory.org/}{https://simonsobservatory.org/}} and the Vera Rubin Observatory \citep{Ivezic19}\footnote{\href{https://www.lsst.org/}{https://www.lsst.org/}}, this trajectory simultaneously entails more stringent accuracy requirements and a subsequently higher number of parameters to describe various systematic effects. At the same time, the $\Lambda$CDM model of cosmology is put under extreme pressure, with ever more complex theoretical models being developed to explain tensions in the values of cosmological parameters, thus growing the size of the parameter space. While these developments promise significant advancements in our understanding of the Universe, they present a formidable challenge for Bayesian inference, stretching the capabilities of traditional inference methods.

We advocate for a new paradigm of cosmological likelihood-based inference to tackle the challenges of next-generation surveys. This approach leverages recent developments in machine learning (ML) and its underlying technology to accelerate both parameter estimation and model selection --- the fundamentals of Bayesian inference --- in high-dimensional settings.
We suggest combining (i) \textit{emulation}; (ii) \textit{differentiable and probabilistic programming}; (iii) \textit{scalable Markov chain Monte Carlo (MCMC) sampling} techniques that exploit gradient information and (iv) \textit{decoupled and scalable Bayesian model selection} techniques computing the Bayesian evidence purely from posterior samples. At the basis of these developments is the ability to exploit modern hardware accelerators, such as graphics processing units (GPUs) and tensor processing units (TPUs), to provide a high degree of parallelization for significant computational acceleration.

Emulation is based on statistical and ML models trained to replicate cosmological quantities of interest with high accuracy in a fraction of the time required by traditional methods. By replacing slow forward models underlying the likelihood function with, e.g. neural networks, which efficiently run on GPUs, accurate physical simulations can be computed with significantly fewer computational resources, providing massive speed-ups.

Additionally, one can pair emulators with a likelihood fully written using differentiable and probabilistic programming languages. This unlocks the acceleration provided by ML environments such as \texttt{TensorFlow} \citep{Abadi15}\footnote{\href{https://www.tensorflow.org/}{https://www.tensorflow.org/}}, \texttt{PyTorch} \citep{Paszke19}\footnote{\href{https://pytorch.org/}{https://pytorch.org/}} and \texttt{JAX} \citep{jax2018github}\footnote{\href{https://github.com/google/jax}{https://github.com/google/jax}} to enable GPU execution and differentiability of the likelihood with respect to the input parameters.  Moreover, probabilistic programming languages can be used, e.g. to easily formulate complex hierarchical models.

Differentiability also unlocks gradient-based sampling algorithms, such as Hamiltonian Monte Carlo \citep[HMC,][]{Duane87, Neal96} and its variants, which represent a promising avenue to address the challenges of high-dimensional inference. These techniques allow one to efficiently explore the complex parameter spaces that characterize cosmological data analyses. Yet, they crucially rely on having access to the derivatives of the likelihood function with respect to input parameters, which are in general expensive and inaccurate, especially when computed by finite differences. This obstacle is being overcome through the development of fully-differentiable frameworks for cosmological data analysis \citep[see e.g.][]{Nygaard23, Ruiz24, Balkenhol24}, which also benefit from the acceleration provided by the same dedicated hardware of ML models.

The complexity of cosmological data analysis is not limited to parameter estimation in high-dimensional settings; it also encompasses the task of model comparison. This is particularly important for cosmology, whose main goal is the identification of the most accurate cosmological model of our Universe given observational data. Bayesian model comparison requires the computation of the model evidence (also known as the marginal likelihood) for different models. Consequently, in order to meaningfully and efficiently compare competing cosmological models in light of new data, it becomes imperative to develop novel methodologies that enable evidence estimation for model comparison in the high-dimensional landscapes that will characterize next-generation surveys. Such methodologies should not be necessarily bound to a particular sampling method in order to provide full flexibility. Being able to compute the evidence from posterior chains independently from the sampling algorithm is thus of paramount importance for enabling next-generation cosmological model comparison.

This paper focuses on both parameter estimation and model comparison in the context of Stage IV likelihood-based cosmological analyses, proposing a comprehensive approach towards cosmological inference from upcoming next-generation datasets. We particularly highlight how parameter estimation and model comparison can be performed from posterior samples obtained using the No U-Turn Sampler \citep[NUTS,][]{Hoffman14}, a highly efficient and adaptive variant of HMC. The expensive cosmological Boltzmann solvers are replaced by the \texttt{CosmoPower-JAX} emulators \citep{CP,CPJ}, and evidence estimation is performed using \texttt{harmonic}, a software implementation of the learned harmonic mean estimator \citep{mcewen2023machine}.

We demonstrate this paradigm on two scenarios. First, we sample the posterior distribution for a simulated Stage IV cosmic shear survey configuration using NUTS and a fully differentiable pipeline. We then compute the evidence from the posterior chains using \texttt{harmonic} and compare the estimate with that obtained using a nested sampler. We perform this operation for two competing cosmological models, with the goal of performing Bayesian model comparison, obtaining results in excellent agreement with nested sampling, but taking only a fraction of the time. We also showcase joint inference on three simulated next-generation surveys, each performing a 3x2pt analysis, obtaining values of the Bayes factor in 8 days, as opposed to the 12 years estimated for the same result to be obtained by traditional techniques.

This paper is structured as follows. We begin in Sect.~\ref{sec:background} with some background on parameter estimation and model comparison, highlighting the challenges associated with the large parameter spaces characterizing upcoming surveys. We then outline in Sect.~\ref{sec:methods} each of the four pillars of the proposed new paradigm of cosmological likelihood-based inference for complete Bayesian analysis, including both parameter estimation and model selection, in high-dimensional settings. In Sect.~\ref{sec:3739} and Sect.~\ref{sec:157159} we introduce two simulated survey scenarios, to validate and demonstrate our framework for Bayesian inference. We conclude in Sect.~\ref{sec:conclusions}.

\section{Background}
Bayesian analyses are the backbone of modern cosmology, providing a principled framework to obtain cosmological parameter constraints and compare models. Here we review the basics of Bayesian inference, discussing both parameter estimation and model comparison.

\label{sec:background}
\subsection{Parameter estimation}
\label{sec:parameter_estimation}
The fundamental goal of Bayesian parameter estimation is to recover an accurate estimate of the posterior distribution $p(\theta|d, \mathcal{M})$, which encapsulates our understanding of the parameters $\theta$ given observed data $d$ and a model $\mathcal{M}$. By employing Bayes' theorem, this distribution is related to the prior distribution $p(\theta|\mathcal{M}) \equiv \pi(\theta)$, the likelihood function $p(d|\theta, \mathcal{M}) \equiv \mathcal{L}(\theta)$, and the model evidence $p(d|\mathcal{M}) \equiv z_{\mathcal{M}}$ through the relation:
\begin{align}
	p(\theta|d, \mathcal{M}) = \frac{p(d|\theta, \mathcal{M})p(\theta|\mathcal{M})}{p(d|\mathcal{M})} = \frac{\mathcal{L}(\theta) \pi(\theta)}{z_{\mathcal{M}}}.
\end{align}

Estimating the posterior distribution through its direct evaluation on a grid of parameters is computationally infeasible due to the curse of dimensionality, which becomes a prohibitive factor even in spaces of moderate dimension. Additionally, the likelihood function often involves solving intricate physical models or running detailed simulations, and the posterior landscape may be multimodal or have complex correlations between parameters. A grid-based evaluation might miss important regions of the parameter space or require an impractically fine grid to capture these features. Sampling algorithms such as MCMC methods, on the other hand, provide a stochastic exploration that can adapt to the shape of the distribution, and allocate computational resources more efficiently by focusing on the regions of parameter space with high posterior density.

Commonly used MCMC algorithms in cosmology include the fast-slow random-walk Metropolis--Hastings \citep{Lewis02, Lewis13}, affine-invariant ensemble \citep{Goodman10, ForemanMackey13}, and nested sampling algorithms \citep{Skilling06, Feroz08, Feroz09, Feroz19, Handley15, Handley15b, Buchner21}, which have found widespread use thanks to their clear formulations and publicly available implementations. Another sampling technique is HMC \citep{Duane87, Neal96}, which is a more efficient algorithm exploiting Hamiltonian dynamics. By leveraging the information coming from the gradients of the likelihood with respect to the input parameters, HMC concentrates sampling in regions of high posterior mass (the so-called typical set), resulting in a more efficient sampler, particularly in high-dimensional settings. A significantly improved version of HMC is NUTS \citep{Hoffman14}, which overcomes the required tuning of hyperparameters for the numerical integration of Hamilton equations by preventing the sampler from taking U-turns. NUTS results in a more efficient exploration of parameter space and faster convergence, and has thus found widespread use.

\subsection{Model comparison}
\label{sec:model_comparison}
For Bayesian model comparison it is necessary to compute the Bayesian evidence
\begin{equation}
	z_\mathcal{M} = \int p(d|\theta, \mathcal{M})p(\theta|\mathcal{M}) \text{d}\theta  \ ,
\end{equation}
which is a challenging computational problem even in moderate dimensional settings.
From the Bayesian evidence it is possible to compute the Bayes factor ($\text{BF}_{12}$), a crucial quantity used to assess which model ($\mathcal{M}_1$ or $\mathcal{M}_2$) is favored given current data $d$ and defined as:
\begin{equation}
	\text{BF}_{12}
	\equiv \frac{p(d|\mathcal{M}_1)}{p(d|\mathcal{M}_2)}
	= \frac{z_{\mathcal{M}_1}}{z_{\mathcal{M}_2}} \ .
	\label{eq:bayes_factor}
\end{equation}
In the common case where the models have equal prior probabilities, i.e. $p(\mathcal{M}_1)=p(\mathcal{M}_2)$, the Bayes factor reduces to the posterior model odds:
\begin{equation}
	\text{BF}_{12}
	= \frac{p(\mathcal{M}_1|d) p(\mathcal{M}_2)}{p(\mathcal{M}_2|d) p(\mathcal{M}_1)}
	= \frac{p(\mathcal{M}_1|d)}{p(\mathcal{M}_2|d)}
	\ .
	\label{eq:bayes_factor_2}
\end{equation}
Estimating the evidence robustly and efficiently can therefore unlock the full potential of Bayesian analyses, in that it allows one to assess which model is preferred by observational data.

Nested sampling, introduced by \citet{Skilling06}, provides a strategy to estimate the evidence for Bayesian model comparison.  By a clever reparameterization of the likelihood in terms of the enclosed prior volume, nested sampling allows the evidence to be computed by a one-dimensional integral.  While nested sampling targets the evidence, as a by-product posterior inferences can also be calculated by appropriate importance weighting \citep{Skilling06}.  While highly successful, nested sampling places tight constraints on how sampling is performed (sampling the prior subject to likelihood isocontour constraints).  Nested sampling couples the sampling strategy to the evidence calculation, limiting flexibility and scalabilty to high-dimensional parameter spaces (a notable exception for high-dimensional inference is proximal nested sampling, although this is only applicable for log-convex likelihoods; \citealt{Cai22}).

Recently, methods to compute the evidence that are agnostic to the sampling technique and only require posterior samples have gained popularity \citep{Dickey71, Trotta07, heavens2017marginal, Jia19, mcewen2023machine, Srinivasan24, Rinaldi24}. Among them, the learned harmonic mean estimator \citep{mcewen2023machine} has been shown to provide robust estimates of the Bayesian evidence in a variety of scenarios, both in likelihood- and simulation-based inference \citep{mcewen2023machine, Polanska23, polanska2024learned, SpurioMancini23}. In particular, \cite{polanska2024learned} integrated normalizing flows into the learned harmonic mean framework to learn the internal importance sampling target distribution, enhancing the robustness and scalability of the estimator.
In contrast to the learned harmonic mean, most alternative approaches
\citep{Srinivasan24, Rinaldi24} compute the evidence for a surrogate of the posterior, limiting their accuracy.
Gaussianized bridge sampling (GBS), introduced by \citet{Jia19}, is similar in spirit to the learned harmonic mean in that requires only posterior samples and considers an internal learned proposal distribution.  In fact, bridge sampling in general can be viewed as a generalization of the standard importance sampling and original harmonic mean approaches to computing the evidence \citep{Gronau17}.  Nevertheless, restricting to the special case of the harmonic mean has numerous advantages. In addition to posterior samples, bridge sampling requires a second set of samples from the proposal distribution. Furthermore, the likelihood must be evaluated at these additional samples, which can add significant computational cost, and the proposal density must also be evaluated at all sample points, both those from the posterior and the proposal. Finally, use of the optimal bridge function \citep{Meng96} necessitates an iterative approach, again adding to the computational cost. The GBS approach suffers from all of these issues, whereas the learned harmonic mean does not. The primary disadvantage of restricting bridge sampling to the case of the harmonic mean is that the internal importance sampling target distribution must have thinner tails than the posterior, which is in any case already solved by the learned harmonic mean approach \citep{mcewen2023machine}.
Another method to compute the Bayes factor is the Savage--Dickey density ratio \citep{Dickey71, Trotta07}, which is however limited to the case of nested models and requires the normalized marginal posterior to be computed.

\section{Methodology}
\label{sec:methods}
We outline the four pillars of a new paradigm of cosmological likelihood-based inference that will be key to successfully tackling the challenges set by Stage IV cosmological surveys, including emulation (typically based on neural networks), differentiable and probabilistic programming, scalable gradient-based MCMC sampling, and scalable evidence estimation that is agnostic to sampling.
All of these pillars provide the ability to exploit modern hardware accelerators, such as GPUs, to provide high degrees of parallelization for significant computational acceleration.

\subsection{Emulation: \normalfont\texttt{CosmoPower-JAX}}
\label{sec:nn_emulation}
The typical bottleneck in a standard Bayesian analysis is the likelihood function, since it incorporates the entire physical knowledge on the observable being probed. Thousands or even millions of likelihood evaluations are needed to explore the posterior distribution in a complete manner, especially in high-dimensional settings. Even in low-dimensional settings, though, the forward model can be computationally expensive since it describes increasingly complex physical models, and extensions to the $\Lambda$CDM model might introduce additional correlations among an increasing number of parameters. Each call to the likelihood function thus has to be fast.

Emulation based on ML techniques, and neural networks in particular, have already shown promising results to accelerate the forward model, while providing excellent agreement with traditional techniques in a fraction of the computational time. Neural networks are trained once on a dataset of key quantities that represent a bottleneck in the forward model, and then can be easily integrated into the likelihood function, providing significant acceleration while retaining the accuracy of the predictions. One additional advantage of neural networks is that they effectively exploit the highly parallel computing provided by modern hardware accelerators, such as GPUs. This allows for faster training, batch prediction and scalable architectures with ever more expressive power.

ML-based emulation has found widespread use in cosmology \citep[see e.g.][]{Auld07, Auld08, ManriqueYus19, Angulo20, Arico21, Nygaard22, Zennaro21, Bonici22, Bonici24}. In this work, we employ the \texttt{CosmoPower-JAX} package \citep{CPJ}\footnote{\href{https://github.com/dpiras/cosmopower-jax}{https://github.com/dpiras/cosmopower-jax}}, a \texttt{JAX} version of the \texttt{CosmoPower} framework \citep{CP}\footnote{\href{https://github.com/alessiospuriomancini/cosmopower}{https://github.com/alessiospuriomancini/cosmopower}}. These dense neural networks were trained to map cosmological parameters to linear and nonlinear matter power spectra, serving as an efficient replacement for the Boltzmann solvers \texttt{CAMB} \citep{Lewis11} or \texttt{CLASS} \citep{Blas2011} to integrate into the likelihood. By replacing Boltzmann solvers with accurate neural networks, we accelerate the likelihood evaluations and thus solve the first significant bottleneck of next-generation surveys.

\subsection{Differentiable and probabilistic programming:\\\normalfont\texttt{JAX} and \normalfont\texttt{NumPyro}}
\label{sec:jax_numpyro}
Within the current ML paradigm, models are trained using optimizers that leverage gradients computed by the backpropagation of gradient information through automatic differentiation \citep{Baydin18}. Consequently, each operation upon which a model may be built must necessarily be differentiable; that is, rules by which tangent and cotangent vectors propagate under the operators' Jacobian must be defined \citep{Wengert64, Rumelhart86, Baydin18}.

Modern ML ecosystems, such as \texttt{JAX} and \texttt{PyTorch}, are constructed with this in mind, requiring that each primitive operation within their libraries is coupled with associated gradient rules. In this way, using the chain rule one can construct more complex operations, which are automatically differentiable \citep{Bartholomew00, Margossian19}.
By automatic differentiation gradients can then be computed efficiently and accurately for complex operations, replacing numerical approximations of the derivatives, which are typically slow and unstable. Primarily due to its relative maturity, \texttt{PyTorch} provides a larger bank of primitive operations. However, \texttt{JAX} provides a more flexible ecosystem that is functional in nature \citep{wadler1992essence}, avoiding boilerplate code that can arise in other ML-oriented frameworks, and being an efficient replacement for \texttt{NumPy} \citep{numpy}; it is thus well suited to the implementation of general physical models.
In \texttt{JAX} complex operations may more easily be expressed, making it a natural ecosystem for the integration of differentiable programming in e.g. cosmological models \citep{Campagne23}, cosmological simulations \citep{Li24} or underlying mathematical methods, such as spherical harmonic transforms \citep{Price24}.

Recently, probabilistic programming languages (PPLs) have been constructed leveraging these differentiable programming ecosystems. Through PPLs, statistical operations such as sampling and conditioning can be expressed as one-line statements, whilst also inheriting the differentiability of the underlying ecosystems upon which they are constructed. For instance, PPLs allows one to easily formulate Bayesian hierarchical models, which are widely used in cosmology \citep[see e.g.][]{Mandel11, Shariff16, Alsing16, Alsing17, Hinton19, Porqueres21, Mandel22, Porqueres23, Loureiro23, Sellentin23, Kostic23, Nguyen24}.

In this work we use \texttt{NumPyro} \citep{Phan19}\footnote{\href{https://num.pyro.ai/en/stable/}{https://num.pyro.ai/en/stable/}}, the \texttt{NumPy} backend for \texttt{Pyro} \citep{Bingham19}. \texttt{NumPyro} is well-documented and easily interfaces with \texttt{JAX}, thus providing a mature framework to build a gradient-based inference pipeline in \texttt{Python}.
We implement the likelihood for the problems considered (see Sect.~\ref{sec:3739} and Sect.~\ref{sec:157159}) in the \texttt{NumPyro} PPL, which requires a separate likelihood implementation.  Critically, the likelihood must be differentiable:  \texttt{CosmoPower-JAX} provides emulation of the matter power spectrum that is differentiable, and we couple it with \texttt{jax-cosmo} \citep{Campagne23}\footnote{\href{https://github.com/DifferentiableUniverseInitiative/jax\_cosmo}{https://github.com/DifferentiableUniverseInitiative/jax\_cosmo}} for differentiable cosmological models to simulate observable spectra.

\subsection{Scalable MCMC sampling: NUTS}
\label{sec:nuts}

HMC is a gradient-based sampling algorithm that is scalable to high-dimensional parameter spaces and does not suffer from the random-walk shortcomings of Metropolis--Hastings algorithms. NUTS further improves on the effective sample size relative to the gradient evaluation of HMC by preventing the sampling trajectory from returning to previously-visited regions of the parameter space. Though it requires a higher number of model evaluations, NUTS typically leads to more reliable convergence, and is thus ideal for sampling in the context of Stage IV surveys.

Traditionally, the gradients required for HMC sampling are computed by finite differences, which is computationally costly and inaccurate.  Since we have a differentiable pipeline, gradients can be computed by automatic differentiation, providing acceleration and improved accuracy. In the future, alternative samplers \citep{Gabrie21, Wong22, Karamanis22a, Karamanis22b}, or new samplers that further improve the sampling efficiency and thus lead to even more robust inference, can be straightforwardly incorporated into this paradigm.

\subsection{Decoupled and scalable Bayesian model comparison: learned harmonic mean}
\label{sec:harmonic}

The learned harmonic mean \citep{mcewen2023machine} provides a robust and scalable estimator of the Bayesian evidence that is decoupled from the sampling strategy. It only requires samples from the posterior, and their corresponding unnormalized probability density. Hence, it can be used with posterior samples obtained through whatever method is best suited for the problem at hand. This property allows the use of the efficient and scalable NUTS sampler, while still being able to estimate the evidence for Bayesian model comparison.

The learned harmonic mean provides an estimator of the reciprocal evidence $\rho=z^{-1}$, defined by
\begin{equation}
	\label{eqn:harmonic_mean_retargeted}
	\hat{\rho} =
	\frac{1}{N} \sum_{i=1}^N
	\frac{\varphi(\theta_i)}{\mathcal{L}(\theta_i) \pi(\theta_i)} ,
	\quad
	\theta_i \sim p(\theta | d),
\end{equation}
where $\varphi(\theta)$ is an arbitrary normalized probability density.
The estimator can be viewed through the lens of importance sampling, with the posterior acting as the sampling density and $\varphi(\theta)$ as the target density \citep[e.g.][]{mcewen2023machine}. If the importance sampling target density has fatter tails than the sampling density, the variance will grow large (even catastrophically) and the estimator will fail. The key to preventing this issue is to find an appropriate target $\varphi(\theta)$ that is normalized and contained within the posterior.  The optimal target density is the normalized posterior itself \citep{mcewen2023machine}, although this requires knowledge of its normalizing constant, which is precisely the quantity we are attempting to estimate.  In the learned harmonic mean the target is learned to approximate the posterior from posterior samples, subject to the constraint that it has narrower tails.  Several benchmark examples have been considered where the learned harmonic mean has been demonstrated to provide precise and accurate evidence estimates \citep{mcewen2023machine}.

More recently, normalizing flows have been integrated within the learned harmonic mean framework for the internal machine learning technique to learn the importance target distribution \citep{Polanska23,polanska2024learned}.  Normalizing flows \citep{papamakarios2021normalizing} can be elegantly coupled with the learned harmonic mean to provide an approach that is more robust, flexible and scalable than the machine learning models considered previously.

Normalizing flows work by taking a simple base distribution, often a standard Gaussian, through a series of invertible transformations with learned parameters. Typically, in order to train the model the Kullback-Leibler divergence between the unknown target and the flow is minimized, resulting in training by maximum likelihood. Using normalizing flows it is possible to approximate potentially complex probability distributions, draw samples from them and evaluate their normalized density.

Once the flow is trained, its probability density can then be concentrated by reducing the ``temperature'' $T$ of its base distribution. Specifically, this is achieved by scaling the base distribution's variance by a factor $T \in (0,1)$. This has the effect of concentrating the base distribution's density, and in turn the trained flow's density due to the continuity and differentiability of the flow \citep{polanska2024learned}. The concentrated flow is then a normalized approximation of the posterior, maintaining a good topological agreement while having thinner tails --- a perfect candidate for the target $\varphi(\theta)$. Estimates of the evidence error and other sanity checks can also be computed \citep{mcewen2023machine,polanska2024learned}. The learned harmonic mean has already been thoroughly validated against nested sampling for numerous cosmological problems \citep{mcewen2023machine, Polanska23, polanska2024learned}, and is implemented in the \texttt{harmonic} open-source \texttt{Python} package\footnote{\href{https://github.com/astro-informatics/harmonic/}{https://github.com/astro-informatics/harmonic/}}. The code is written in \texttt{JAX} \citep{jax2018github} to provide automatic differentiation and an efficient and scalable implementation that can be run on accelerators such as GPUs.

Critically, this approach decouples sampling from evidence calculation, allowing the use of scalable MCMC sampling techniques (e.g. NUTS).  By integrating normalizing flows inside the overarching statistical framework of the learned harmonic mean we recover a robust estimator of the evidence that is also computationally scalable, proving the missing final component of a complete Bayesian analysis.

\begin{table*}
	\caption{Evidence and Bayes factors for the two analyses considered in this work. The evidence values obtained with nested sampling are the mean and standard deviation across three independent runs with different seeds, while the computation time is the typical one for a single run, and includes sampling and evidence estimation. The likelihood functions sampled with nested sampling and the No U-Turn Sampler (NUTS) are implemented using different packages and include different normalizations; despite these differences, we recover consistent values of the Bayes factor (BF) between $\Lambda$CDM and $w_0w_a$CDM.}
	\label{table:results}
	\centering
	\subfloat[Cosmic shear with 37 ($\Lambda$CDM) and 39 ($w_0w_a\text{CDM}$) parameters, described in Sect.~\ref{sec:3739}.]{%
		\def\arraystretch{1.7}
		\addtolength{\tabcolsep}{-0.25em}
		\begin{tabular}{cccccc}  \toprule
			Method                                                                                                      & $\log(z_{\Lambda\text{CDM}})$ & $\log(z_{w_0w_a\text{CDM}})$ & $\log\text{BF}$ & Total computation time                                                                                                                           \\ \midrule
			\texttt{CAMB} + nested sampling                                                                             & $-107.03 \pm 0.27$            & $-107.81 \pm 0.74$           & $0.78 \pm 0.79$ & $\sim$8 months (48 CPUs)                                                                                                                         \\
			\texttt{CosmoPower-JAX} + NUTS + \texttt{harmonic}                                                          & $40956.55 \pm 0.06$           & $40955.03\pm0.04$            & $1.53 \pm 0.07$ & \def\arraystretch{1}\begin{tabular}{c}2 days (sampling, 12 GPUs) $+$\\12 minutes (evidence, 1 GPU + 48 CPUs) \end{tabular}\def\arraystretch{1.7} \\
			\texttt{CosmoPower-JAX} + NUTS + \def\arraystretch{1}\begin{tabular}{c}na\"ive flow\\estimator\end{tabular} & $400958 \pm 5$                & $40957 \pm 4$                & $1 \pm 6$       & Similar to \texttt{harmonic}                                                                                                                     \\
			\bottomrule
		\end{tabular}}
	\\ \vspace{0.5cm}
	\subfloat[3x(3x2pt) with 157 ($\Lambda$CDM) and 159 ($w_0w_a\text{CDM}$) parameters, described in Sect.~\ref{sec:157159}.]{%
		\def\arraystretch{1.7}
		\addtolength{\tabcolsep}{-0.2em}
		\begin{tabular}{cccccc}  \toprule
			Method                                                                                                      & $\log(z_{\Lambda\text{CDM}})$ & $\log(z_{w_0w_a\text{CDM}})$ & $\log\text{BF}$     & Total computation time                                                                                                                           \\ \midrule
			\texttt{CAMB} + nested sampling                                                                             & Unfeasible                    & Unfeasible                   & Unfeasible          & 12 years (projected, 48 CPUs)                                                                                                                    \\
			\texttt{CosmoPower-JAX} + NUTS + \texttt{harmonic}                                                          & $406689.6^{+0.5}_{-0.3}$      & $406687.7^{+0.5}_{-0.3}$     & $1.9^{+0.7}_{-0.5}$ & \def\arraystretch{1}\begin{tabular}{c}8 days (sampling, 24 GPUs) $+$\\17 minutes (evidence, 1 GPU + 48 CPUs) \end{tabular}\def\arraystretch{1.7} \\
			\texttt{CosmoPower-JAX} + NUTS + \def\arraystretch{1}\begin{tabular}{c}na\"ive flow\\estimator\end{tabular} & $406703 \pm 39$               & $406701 \pm 62$              & $2\pm 73$           & Similar to \texttt{harmonic}                                                                                                                     \\
			\bottomrule
		\end{tabular}}
\end{table*}

\begin{figure*}
	\centering
	\includegraphics[width=\linewidth]{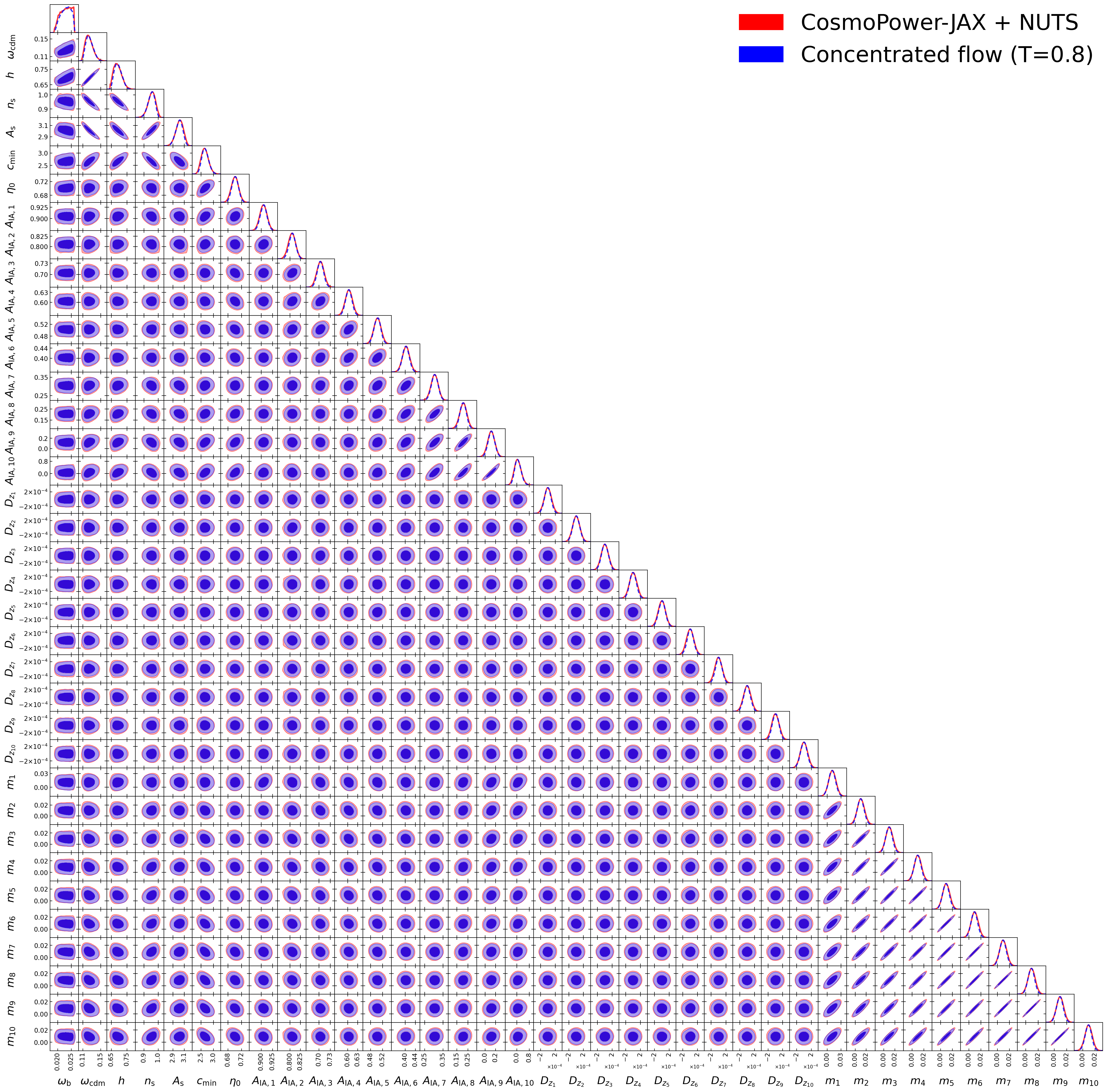}
	\caption{Corner plot for the $37$-dimensional $\Lambda$CDM model, showing the posterior contours obtained with \texttt{CosmoPower-JAX} in red and the concentrated flow with temperature $T=0.8$ in blue. The $A_{\mathrm{s}}$ parameter indicates $\ln\left( 10^{10} A_{\mathrm{s}} \right)$.}
	\label{fig:37_lcdm_corner}
\end{figure*}

\begin{figure*}
	\centering
	\includegraphics[width=\linewidth]{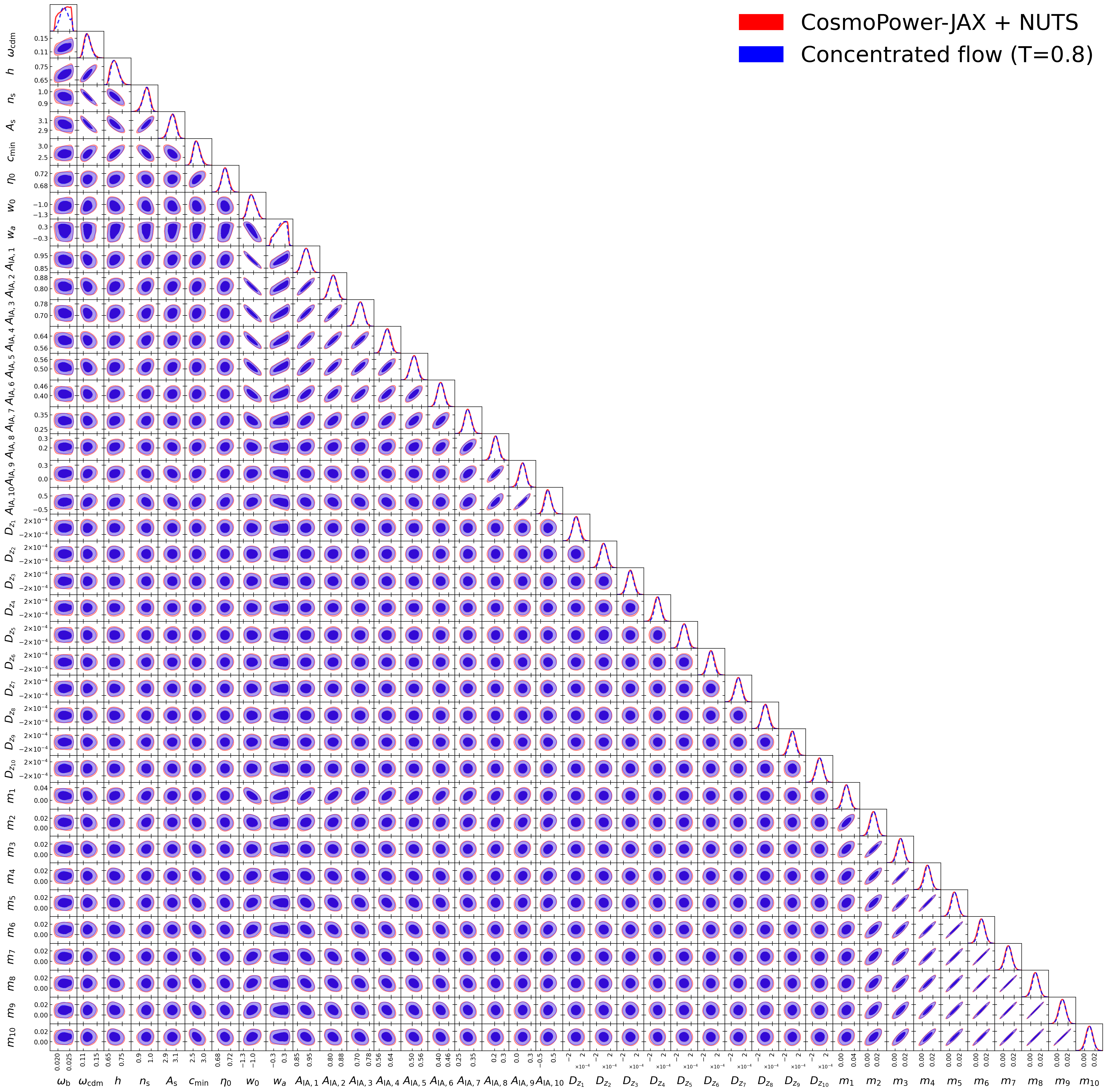}
	\caption{Same as Fig.~\ref{fig:37_lcdm_corner} for the $39$-dimensional $w_{0}w_{a}$CDM model.}
	\label{fig:39_wcdm_corner}
\end{figure*}

\section{Cosmic shear analysis\\with 37 and 39 parameters}
\label{sec:3739}

We first demonstrate the application of the proposed paradigm for a simulated next-generation survey performing a power spectrum cosmic shear analysis. Our goal is to estimate the Bayes factor between the $\Lambda$CDM and $w_0w_a$CDM models in high-dimensional scenarios, exposing the challenges of next-generation surveys if approached with traditional methods, while validating and showcasing the scalability of our approach.

\subsection{Likelihood}

The likelihood details are the same as in \citet{CPJ}, which we summarize here, following the notation of \citet{CP}.
We assume a tomographic survey with $N_{\text{bins}}=10$ bins, where each galaxy is placed in a bin according to its estimated photometric redshift $z$. We compute the angular power spectra between pairs of redshift bins $i, j = 1, \dots , N_{\text{bins}}$ and for different probes. The cosmic shear angular power spectrum $C^{\epsilon \epsilon}_{i j} (\ell)$ is defined as:
\begin{equation}
	C^{\epsilon \epsilon}_{i j} (\ell) = C^{\gamma \gamma}_{i j} (\ell) + C^{\gamma I}_{i j} (\ell) + C^{I \gamma}_{i j} (\ell) + C^{I I}_{i j} (\ell) \ ,
	\label{eq:cepseps}
\end{equation}
which includes the contributions from pure shear ($\gamma$) and intrinsic alignment ($I$). Assuming the extended Limber approximation \citep{LoVerde08}, we can write:
\begin{equation}
	C^{A B}_{i j} (\ell) = \int_0^{\chi_H} \frac{W_i^A (\chi)  W_j^B (\chi)}{\chi^2}  P_{\delta \delta} \left(k = \frac{\ell + 1/2}{\chi}, z \right) \mathrm{d}\chi \ ,
	\label{eq:cgeneric}
\end{equation}
where $P_{\delta\delta}(k, z)$ is the matter power spectrum, $\chi_H = c/H_0$ (with $c$ the speed of light and $H_0$ the Hubble constant today) is the Hubble radius, and $W (\chi)$ indicates a window function for each probe $\{A, B\} = \{\gamma, I \}$ as a function of the comoving distance $\chi$.

For the pure cosmic shear $\gamma$ the window function can be written as:
\begin{equation}
	\label{eq:window_shear}
	W_i^\gamma (\chi) = \frac{3 \Omega_{\rm m} H_0^2 }{2  c^2} \frac{\chi}{a} \int_{\chi}^{\chi_{\rm H}}   n_{i, \textrm{source}}(\chi')  \frac{\chi'-\chi}{\chi'} \mathrm{d} \chi' \ ,
\end{equation}
where $a$ is the scale factor, $\Omega_{\rm m}$ is the matter density parameter today and $n_{i,\text{source}}(z)$ represents the tomographic redshift bin distribution of the sources being observed. For the intrinsic alignment field, we consider the non-linear alignment model (NLA, \citealp{Hirata04, Joachimi11}) modified as in \citet{CPJ}:
\begin{equation}
	\label{eq:window_IA}
	W_i^{\rm I}(\chi) = - A_{\mathrm{IA}, i}  \frac{C_1  \rho_{\rm cr}  \Omega_{\rm m}}{D(\chi)}  n_{i, \textrm{source}}(\chi) \ ,
\end{equation}
where $D(\chi)$ is the linear growth factor, $\rho_{\rm cr}$ is the critical density of the Universe, $C_1$ is a constant, and with one intrinsic alignment amplitude for each redshift bin $A_{\mathrm{IA}, i}$, to allow for more flexibility in the modeling. For each redshift bin we also include a multiplicative bias $m_i$ \citep{Huterer06, Amara08, Kitching15, Taylor18, Mandelbaum18}, which rescales the cosmic shear power spectrum by a factor $(1 + m_i)(1 + m_j )$, and a shift parameter $D_{z_i,\text{source}}$ \citep{Eifler21}, which shifts the mean of the bin redshift distribution so that we actually consider $n_{i, \text{source}}'(z) = n_{i, \text{source}}(z-D_{z_i, \text{source}})$.

Finally, we model the redshift distributions with kernel density estimation (KDE), and consider a Gaussian likelihood with a simulated covariance matrix as in \cite{Tutusaus20}, with surface density of galaxies $n_{\text{source}} = 30$ galaxies/arcmin$^2$, observed ellipticity dispersion $\sigma_{\epsilon} = 0.3$, and sky fraction $f_\text{sky} = 0.35$. We compute each $C(\ell)$ spectrum for 30 log-spaced bin values between $\ell_{\text{min}} = 30$ and $\ell_{\text{max}} = 3000$. To compute the theoretical predictions for the cosmological observables we use the Core Cosmology Library (\texttt{CCL}, \citealp{Chisari19}).

\subsection{Models}

We compare two cosmological models, namely the fiducial $\Lambda$CDM model (considered also in \citealt{CPJ}) and the $w_0w_a$CDM model, where in the latter we assume that the dark energy equation of state evolves with cosmic time according to:
\begin{equation}
	w(a) = w_0 + (1-a) w_a \ ,
	\label{eq:w0wa}
\end{equation}
which is the most common parametrization (also known as the CPL parametrization) for dynamic dark energy \citep{Chevallier01, Linder03}. The $w_0w_a$CDM model introduces two extra parameters ($w_0$ and $w_a$), and $\Lambda$CDM is recovered for $w_0=-1$ and $w_a=0$. The simulated data vector is generated assuming a $\Lambda$CDM model, so we expect that the Bayes factor should favor this hypothesis. The prior distributions on the cosmological and nuisance parameters are the same as those in \citet{CPJ}, with the addition of uniform distributions for $w_0$ and $w_a$ between $(-1.5, -0.5)$ and $(-0.5, 0.5)$, respectively. The total number of parameters being sampled is thus 37 and 39 for $\Lambda$CDM and $w_0w_a$CDM, respectively.

\subsection{Computational approaches}

We compare three approaches to perform parameter estimation and model comparison: the ``traditional'' approach, the proposed ``future'' paradigm, and an approach based on a na\"ive flow estimation of the evidence for comparison \citep[discussed in][]{SpurioMancini23, polanska2024learned}.

First, we use the nested sampler \texttt{PolyChord} \citep{Handley15,Handley15b} to sample the posterior distribution and run the inference pipeline within Cobaya \citep{Torrado21}, using \texttt{CAMB} to predict the matter power spectrum. We run \texttt{PolyChord} with default settings on 48 CPU cores twice, each time assuming either $\Lambda$CDM or $w_0w_a$CDM. Since we found the values of the evidence to fluctuate significantly between different runs, we repeated the process three times for each model, quoting the mean and standard deviation of the $\log$ evidence, and the average time to obtain the Bayes factor.

Second, we compare this ``traditional`` approach with our ``future'' paradigm, which replaces \texttt{CAMB} with the \texttt{CosmoPower-JAX} emulator, rewrites the likelihood in the auto-differentiable language \texttt{JAX}, making use of \texttt{jax-cosmo}, samples the posterior distribution with NUTS implemented in \texttt{NumPyro}, accelerated by automatic differentiation, and computes a robust estimate of the evidence from posterior samples using the learned harmonic mean implemented in \texttt{harmonic}.
We run the NUTS chains in parallel on 12 A100 80GB GPUs, collecting 60 chains of 2000 samples each (after 400 samples of warm-up) for both $\Lambda$CDM and $w_0w_a$CDM. As in \cite{CPJ}, parallelization is obtained by running multiple chains on each GPU.
For \texttt{harmonic}, we train a rational-quadratic spline flow \citep{durkan2019neural}, consisting of 4 layers and 64 spline bins, on a single GPU. We separate the NUTS samples into a training set the flow is trained on and an inference set that is used for estimation. We use $30\%$ of chains for training and the remaining $70\%$ for inference to compute the evidence on 48 CPUs with temperature $T=0.8$.

Third, the evidence can also be estimated directly from the normalizing flow in an approach introduced by \citet{SpurioMancini23}, and further discussed in \citet{polanska2024learned},  called the ``na\"ive flow estimator''.  Since the flow is a normalized approximation of the posterior, the expectation of the ratio of the unnormalized posterior density and the flow density should be the evidence itself. This can be estimated via a Monte Carlo expectation as the mean of the ratios evaluated across samples from the posterior. In contrast to the learned harmonic mean, this approach is highly sensitive to  the flow being a close approximation of the posterior and consequently can suffer from a large bias and variance. We also consider this method of estimating the evidence to demonstrate the need for an alternative principled estimator of the evidence, as provided by the learned harmonic mean.  We compare the results obtained replacing \texttt{harmonic} with the na\"ive flow estimator on the same NUTS samples and hardware.

\subsection{Results}
\label{sec:results_3739}
The results are reported in Table~\ref{table:results}(a). The values of the evidence when using \texttt{CAMB} $+$ nested sampling and \texttt{CosmoPower-JAX} $+$ NUTS $+$ \texttt{harmonic} use different implementations of the likelihood (the latter implemented in a probabilistic programming framework), including different normalizations, so their values cannot be compared directly. Nevertheless, the $\log$ BFs are in good agreement between the different approaches, with $0.78\pm0.79$ for nested sampling and $1.53\pm0.07$ for \texttt{harmonic}.  Note that the evidence values correctly favour the ground truth model considered for the simulated data.  While the evidence computed by the na\"ive flow estimator is also in agreement with a value of $1\pm6$, notice that its error is considerably larger than the other estimators, as anticipated.  While the evidence values computed by the ``traditional'' and ``future'' paradigms are in close agreement, computational times are dramatically different, with the ``future'' paradigm being two orders of magnitude faster, taking 2 days (with only 12 additional minutes required to compute the evidence) instead of roughly 8 months. Note that with respect to the analysis made by \citet{CPJ}, the computation times that we report here are higher since they include two models ($\Lambda$CDM and $w_0w_a$CDM), as well as more chains to obtain a robust estimate of the evidence from \texttt{harmonic}. We also tested that running \texttt{PolyChord} in combination with the \texttt{CosmoPower} emulator yields consistent results, while taking about 16 days on 48 CPU cores (namely 2 runs of 8 days each to estimate the Bayes factor). We additionally report that a single evaluation of the log-likelihood gradient in forward (reverse) mode with our approach takes about 335 (80) seconds on a single CPU, and 15 (20) seconds on a single GPU, demonstrating the acceleration provided by modern hardware.

The contours for the posterior samples (red) alongside the concentrated flow at $T=0.8$ (blue) used for inference for this analysis are shown in Fig.~\ref{fig:37_lcdm_corner} and Fig.~\ref{fig:39_wcdm_corner} for the $\Lambda$CDM and $w_0w_a$CDM models respectively. A requirement for the learned harmonic mean estimator is that the concentrated flow is contained within the posterior. While it can be informative to inspect corner plots to ensure that, it is useful to consider several additional sanity checks, especially in high-dimensional contexts. In particular, we confirm that the values of the error estimate, kurtosis and the ratio between the square root variance of variance and variance estimates, introduced in detail by \citet{mcewen2023machine}, are in agreement with the theoretical expectations.

\section{3x2pt analysis with 157 and 159 parameters}
\label{sec:157159}

We further showcase the scalability and robustness of our proposed paradigm by considering three different next-generation simulated surveys, each performing a 3x2pt analysis, where information on cosmic shear, galaxy clustering and their cross-correlation is combined to obtain more stringent constraints on the cosmological parameters \citep{Joachimi10}. We refer to this application as a 3x(3x2pt) analysis.
We present the likelihood and results below; the cosmological models and computational approaches are the same as Sect.~\ref{sec:3739}, although as we discuss in Sect.~\ref{sec:results_157159} we are not able to run the ``traditional'' approach in this high-dimensional setting.

\subsection{Likelihood}
The galaxy clustering field (n) power spectrum $C^{\text{nn}}_{ij} (\ell)$ can be expressed using Eq.~(\ref{eq:cgeneric}) assuming that galaxies are a linearly-biased tracer of dark matter, with a free parameter $b_i$ for each redshift bin. The corresponding window function can then be written as:
\begin{equation}
	W_i^{\text{n}}(\chi) = b_i \, n_{i,\text{lens}}(\chi) \ ,
	\label{eq:window_gc}
\end{equation}
where $n_{i,\text{lens}}$ is a different sample of redshift distributions, also including a shift for each bin $D_{z_i,\text{lens}}$. Finally, the cross power spectrum between the shear field and the galaxy clustering field, also called the galaxy-galaxy lensing power spectrum, is written as:
\begin{equation}
	C_{ij}^{\text{n} \epsilon}(\ell) = C_{ij}^{\text{n} \gamma}(\ell) + C_{ij}^{\text{n} \text{I}}(\ell) \ .
	\label{eq:cell_cross}
\end{equation}
The total number of parameters for this cross-survey analysis is 157 for $\Lambda$CDM and 159 for $w_0w_a$CDM, since each survey comes with its own 50 nuisance parameters, and all surveys share 7 (or 9) cosmological parameters. The specific details of every survey are reported in \citet{CPJ}.

\subsection{Results}
\label{sec:results_157159}
All numerical results for this higher dimensional analysis are reported in Table~\ref{table:results}(b).
The high number of parameters and the need to consider multiple probes pose a significant challenge for ``traditional'' methods, despite being quite realistic for next-generation surveys. We estimated that the computation of the Bayes factor in such a scenario would require about 12 years on 48 CPU cores, making it effectively impossible to perform model selection in a reasonable time. While an emulator could speed up nested sampling, it is unlikely that it could scale the analysis to such high-dimensional scenarios, given the relatively large variance of the ``traditional'' method already shown in Sect.~\ref{sec:3739}. In our experiments, we found that even the 3x2pt analysis of a single survey with 57 parameters combining an emulator with \texttt{PolyChord} (with default hyperparameters) would struggle to converge in a reasonable time for the likelihood considered here.

On the other hand, the proposed ``future'' paradigm provides an estimate of the Bayes factor based on 40 (45) converged chains --- with 1500 samples each, after 1000 samples of warm-up --- for the $\Lambda$CDM ($w_0w_a$CDM) model, totalling 8 days on 24 GPUs, with only 17 additional minutes required to compute the evidence. In this case, the flow model we consider is constructed from 2 layers and 128 spline bins. The $\log$ Bayes factor computed by \texttt{harmonic} is $1.9^{+0.7}_{-0.5}$, which correctly favours the ground truth model considered for the simulated data. The evidence computed by the na\"ive flow estimator is $2\pm73$: its error is very large, as anticipated, rendering this approach unusable in practice. Again, the computation times reported here are higher than what presented in \citet{CPJ} since we are now performing model comparison, and a higher number of chains was required to satisfy the diagnostics provided by \texttt{harmonic}.

\begin{figure}
	\centering
	\includegraphics[width=\linewidth]{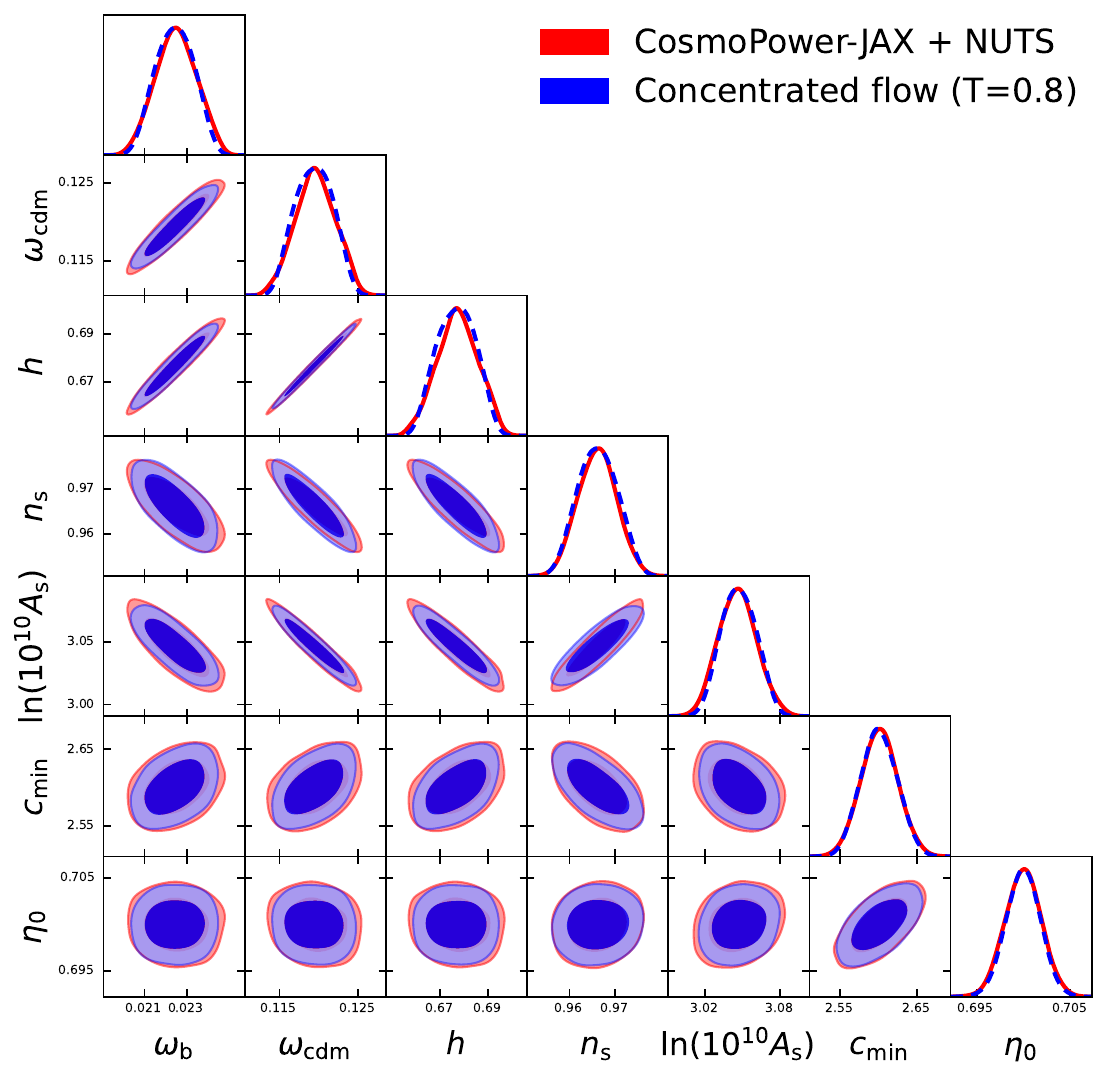}
	\caption{Corner plot of the subset of cosmological parameters for the $157$-dimensional $\Lambda$CDM model, showing the posterior contours in red and the concentrated flow with temperature $T=0.8$ in blue.}
	\label{fig:157_lcdm_cosmo}
	\vspace{0.05cm}
\end{figure}

\begin{figure}
	\centering
	\includegraphics[width=\linewidth]{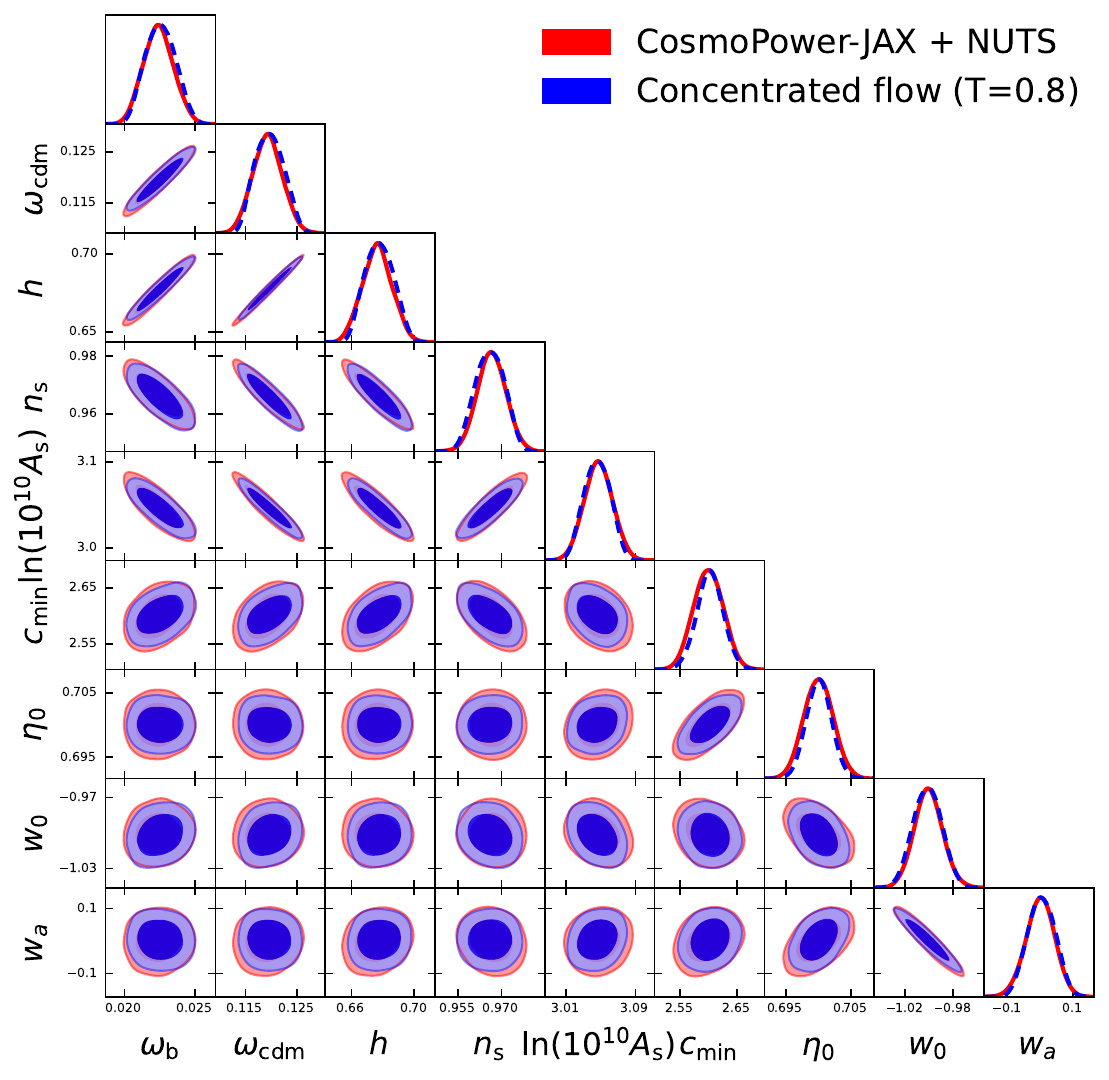}
	\caption{Same as Fig.~\ref{fig:157_lcdm_cosmo} for the $159$-dimensional $w_{0}w_{a}$CDM model.}
	\label{fig:159_wcdm_cosmo}
\end{figure}

\begin{figure*}
	\centering
	\includegraphics[width=0.85\linewidth]{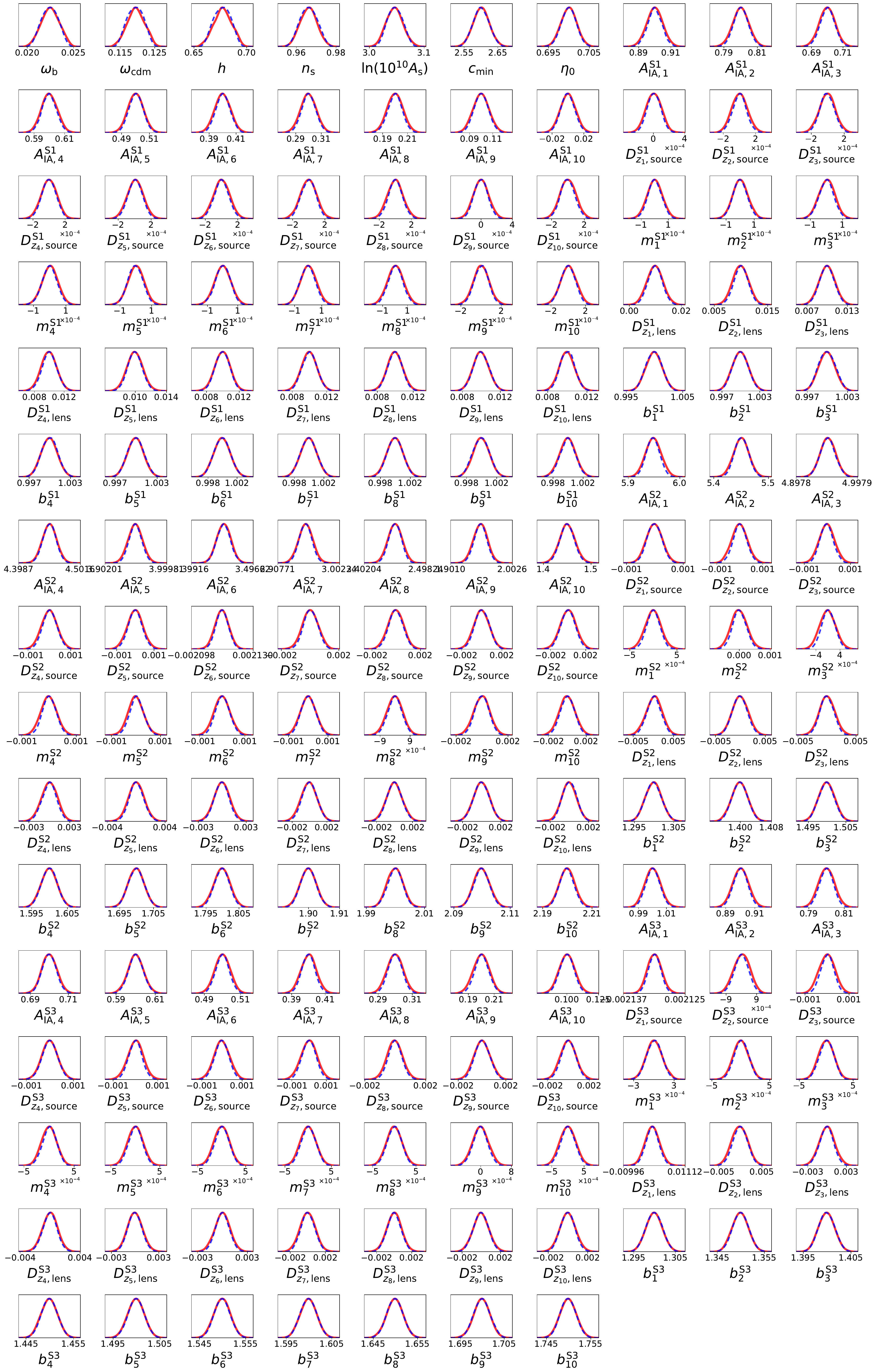}
	\caption{Marginal distributions of all parameters for the $157$-dimensional $\Lambda$CDM model, with the posterior distribution obtained with \texttt{CosmoPower-JAX} in red, and the concentrated flow with temperature $T=0.8$ in blue.}
	\label{fig:157_lcdm_marginals}
\end{figure*}

\begin{figure*}
	\centering
	\includegraphics[width=0.9\linewidth]{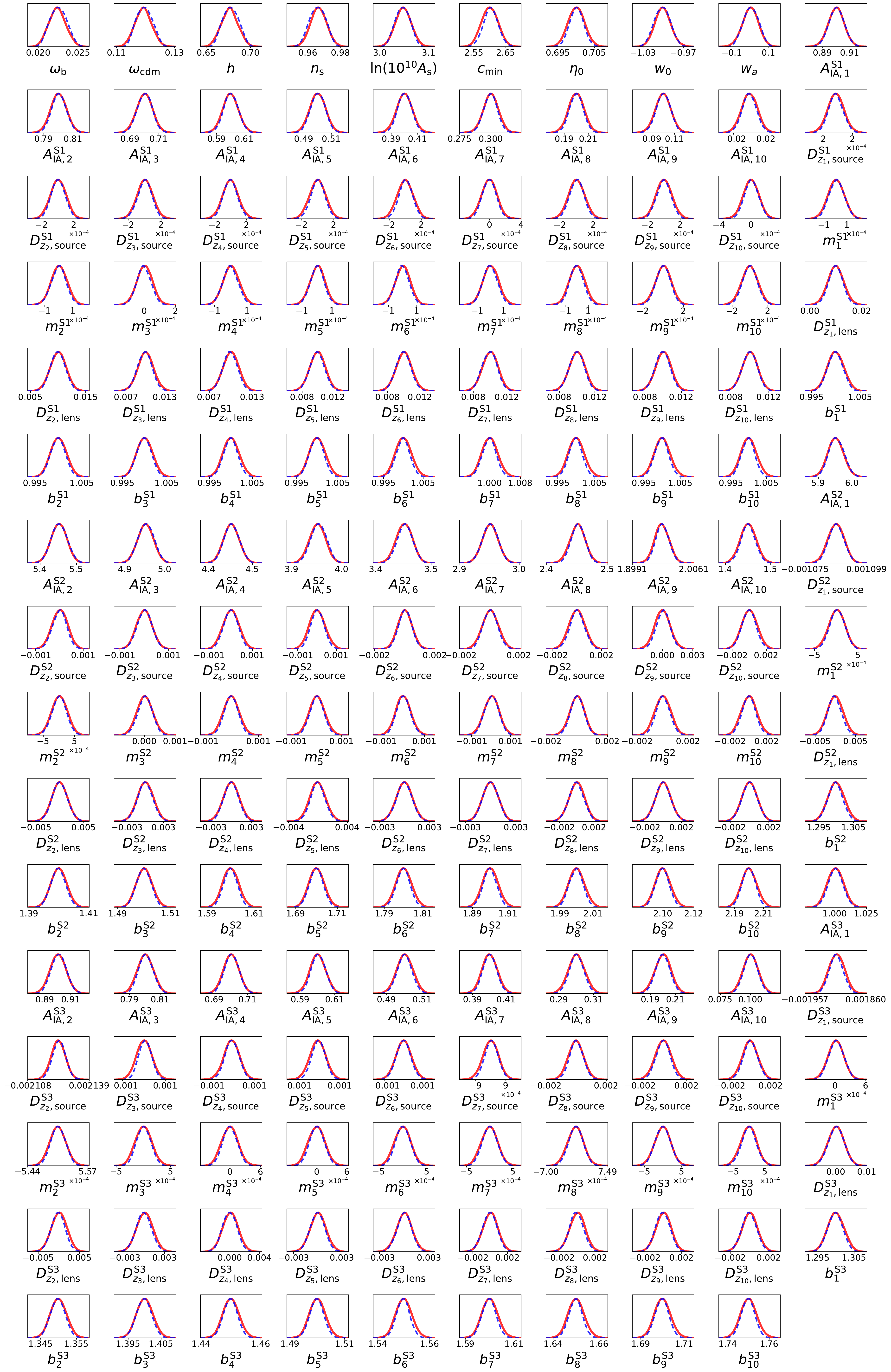}
	\caption{Same as Fig.~\ref{fig:157_lcdm_marginals} for the $159$-dimensional $w_{0}w_{a}$CDM model.}
	\label{fig:159_wcdm_marginals}
\end{figure*}

The contours plots for the posterior samples (red) alongside the concentrated flow at $T=0.8$ (blue) used for inference for this analysis are shown in Fig.~\ref{fig:157_lcdm_cosmo} and Fig.~\ref{fig:159_wcdm_cosmo} for the $\Lambda$CDM and $w_0w_a$CDM models respectively. Fig.~\ref{fig:157_lcdm_marginals} and Fig.~\ref{fig:159_wcdm_marginals} show the one-dimensional marginal plots of the posterior samples (red) alongside the concentrated flow at $T=0.8$ (blue), for all the model parameters for $\Lambda$CDM and $w_0w_a$CDM respectively. As in Sect.~\ref{sec:results_3739}, we ensure that we satisfy the sanity checks introduced by \citet{mcewen2023machine}.

\section{Conclusions}
\label{sec:conclusions}
We propose a combination of state-of-the-art techniques from machine learning (ML), differentiable and probabilistic programming, high-dimensional sampling and robust statistical estimation to tackle the Bayesian likelihood-based cosmological analyses of the future. Instead of relying on nested sampling with standard Boltzmann solvers, we leverage ML-based emulation (\texttt{CosmoPower-JAX}), differentiable programming (\texttt{JAX}), probabilistic programming (\texttt{NumPyro}), more efficient samplers (NUTS), and robust evidence estimation (\texttt{harmonic}) to perform end-to-end cosmological analyses including parameter estimation and model comparison in two challenging cosmological scenarios scaling up to 159 parameters.

We first computed the Bayes factor for a simulated Stage IV cosmic shear analysis comparing the $\Lambda$CDM model with 37 parameters and a dynamical dark energy model $w_0w_a$CDM with 39 parameters. We demonstrated excellent agreement with nested sampling despite different implementations of the likelihood, while requiring two orders of magnitude less computational time. We additionally demonstrated the scalability of our approach by considering the joint analysis of three Stage IV surveys, each performing a 3x2pt analysis, totalling either 157 ($\Lambda$CDM) or 159 ($w_0w_a$CDM) parameters. We computed the Bayes factor obtaining results requiring only 8 days on 24 GPUs, while projecting that a ``traditional" analysis on 48 CPU cores would require 12 years, assuming nested sampling is capable of scaling to such a high-dimensional setting.

We advocate for a combination of ML emulation, differentiable and probabilistic programming, scalable sampling and robust evidence estimation to tackle forthcoming cosmological likelihood-based analyses. This ``future" paradigm unlocks parameter estimation and model comparison for Stage IV cosmological surveys with an unprecedented number of parameters. Given that all packages used in our analysis are already publicly available, we envision this approach could become the standard in future likelihood-based analyses, allowing the community to fully analyze the data derived from upcoming observations in a practical timescale.

\section*{Author contributions}
\textbf{DP}: conceptualization; formal analysis; investigation; methodology; validation; software; visualization; resources; writing -- original draft.
\textbf{AP}: formal analysis; methodology; validation; software; visualization; writing -- original draft.
\textbf{ASM}: conceptualization; methodology; validation; software; supervision; writing -- review \& editing.
\textbf{MAP}: software; valididation; writing -- review \& editing.
\textbf{JDM}: conceptualization; methodology; software; validation; project administration; funding acquisition; supervision; writing -- review \& editing.

\section*{Acknowledgements}
DP was supported by a Swiss National Science Foundation (SNSF) Professorship grant (No. 202671), and by the SNF Sinergia grant CRSII5-193826 “AstroSignals: A New Window on the Universe, with the New Generation of Large Radio-Astronomy Facilities”. AP is supported by the UCL Centre for Doctoral Training in Data Intensive Science (STFC grant number ST/W00674X/1). ASM acknowledges support from the MSSL STFC Consolidated Grant ST/W001136/1. MAP and JDM are supported in part by EPSRC (grant number EP/W007673/1). The authors are pleased to acknowledge that part of the work reported on in this paper was performed using the Princeton Research Computing resources at Princeton University which is a consortium of groups led by the Princeton Institute for Computational Science and Engineering (PICSciE) and Office of Information Technology’s Research Computing.

\bibliographystyle{mnras}
\bibliography{paper}
\end{document}